\documentclass[pra,aps,showpacs,twocolumn,superscriptaddress]{revtex4}
\usepackage{amsmath,amsfonts,amssymb,caption,hyperref,color,epsfig,graphics,subfigure,graphicx,latexsym,mathrsfs,revsymb,theorem,url,verbatim,epstopdf}
\allowdisplaybreaks[3]

\hypersetup{colorlinks,linkcolor={blue},citecolor={blue},urlcolor={red}}

\usepackage{hyperref}

\makeatletter

\newcommand{\Rmnum}[1]{\expandafter\@slowromancap\romannumeral #1@}
\makeatother
\newtheorem{definition}{Definition}
\newtheorem{proposition}[definition]{Proposition}

\newtheorem{Theorem}[definition]{Theorem}

\newtheorem{conjecture}[definition]{Conjecture}

\newtheorem{remark}[definition]{Remark}
\newtheorem{example}[definition]{Example}
\newtheorem{question}[definition]{Question}

\def\squareforqed{\hbox{\rlap{$\sqcap$}$\sqcup$}}
\def\qed{\ifmmode\squareforqed\else{\unskip\nobreak\hfil
		\penalty50\hskip1em\null\nobreak\hfil\squareforqed
		\parfillskip=0pt\finalhyphendemerits=0\endgraf}\fi}
\def\endenv{\ifmmode\;\else{\unskip\nobreak\hfil
		\penalty50\hskip1em\null\nobreak\hfil\;
		\parfillskip=0pt\finalhyphendemerits=0\endgraf}\fi}
\newenvironment{proof}{\noindent \textbf{{Proof.~} }}{\qed}
\def\Dbar{\leavevmode\lower.6ex\hbox to 0pt
	{\hskip-.23ex\accent"16\hss}D}
\makeatletter
\def\url@leostyle{%
	\@ifundefined{selectfont}{\def\UrlFont{\sf}}{\def\UrlFont{\small\ttfamily}}}
\makeatother
\def\bcj{\begin{conjecture}}
	\def\ecj{\end{conjecture}}
\def\bcr{\begin{corollary}}
	\def\ecr{\end{corollary}}
\def\bd{\begin{definition}}
	\def\ed{\end{definition}}
\def\bea{\begin{eqnarray}}
\def\eea{\end{eqnarray}}
\def\bem{\begin{enumerate}}
	\def\eem{\end{enumerate}}
\def\bex{\begin{example}}
	\def\eex{\end{example}}
\def\bim{\begin{itemize}}
	\def\eim{\end{itemize}}
\def\bl{\begin{lemma}}
	\def\el{\end{lemma}}
\def\bma{\begin{bmatrix}}
	\def\ema{\end{bmatrix}}
\def\bpf{\begin{proof}}
	\def\epf{\end{proof}}
\def\bpp{\begin{proposition}}
	\def\epp{\end{proposition}}
\def\bqu{\begin{question}}
	\def\equ{\end{question}}
\def\br{\begin{remark}}
	\def\er{\end{remark}}
\def\bt{\begin{theorem}}
	\def\et{\end{theorem}}

\def\btb{\begin{tabular}}
	\def\etb{\end{tabular}}

\newcommand{\nc}{\newcommand}

\def\a{\alpha}

\def\e{\epsilon}

\def\r{\rho}
\def\s{\sigma}

\nc{\bbA}{\mathbb{A}} \nc{\bbB}{\mathbb{B}} \nc{\bbC}{\mathbb{C}}
\nc{\bbD}{\mathbb{D}} \nc{\bbE}{\mathbb{E}} \nc{\bbF}{\mathbb{F}}
\nc{\bbG}{\mathbb{G}} \nc{\bbH}{\mathbb{H}} \nc{\bbI}{\mathbb{I}}
\nc{\bbJ}{\mathbb{J}} \nc{\bbK}{\mathbb{K}} \nc{\bbL}{\mathbb{L}}
\nc{\bbM}{\mathbb{M}} \nc{\bbN}{\mathbb{N}} \nc{\bbO}{\mathbb{O}}
\nc{\bbP}{\mathbb{P}} \nc{\bbQ}{\mathbb{Q}} \nc{\bbR}{\mathbb{R}}
\nc{\bbS}{\mathbb{S}} \nc{\bbT}{\mathbb{T}} \nc{\bbU}{\mathbb{U}}
\nc{\bbV}{\mathbb{V}} \nc{\bbW}{\mathbb{W}} \nc{\bbX}{\mathbb{X}}
\nc{\bbZ}{\mathbb{Z}}

\nc{\bA}{{\bf A}} \nc{\bB}{{\bf B}} \nc{\bC}{{\bf C}}
\nc{\bD}{{\bf D}} \nc{\bE}{{\bf E}} \nc{\bF}{{\bf F}}
\nc{\bG}{{\bf G}} \nc{\bH}{{\bf H}} \nc{\bI}{{\bf I}}
\nc{\bJ}{{\bf J}} \nc{\bK}{{\bf K}} \nc{\bL}{{\bf L}}
\nc{\bM}{{\bf M}} \nc{\bN}{{\bf N}} \nc{\bO}{{\bf O}}
\nc{\bP}{{\bf P}} \nc{\bQ}{{\bf Q}} \nc{\bR}{{\bf R}}
\nc{\bS}{{\bf S}} \nc{\bT}{{\bf T}} \nc{\bU}{{\bf U}}
\nc{\bV}{{\bf V}} \nc{\bW}{{\bf W}} \nc{\bX}{{\bf X}}
\nc{\bZ}{{\bf Z}}

\nc{\cA}{{\cal A}} \nc{\cB}{{\cal B}} \nc{\cC}{{\cal C}}
\nc{\cD}{{\cal D}} \nc{\cE}{{\cal E}} \nc{\cF}{{\cal F}}
\nc{\cG}{{\cal G}} \nc{\cH}{{\cal H}} \nc{\cI}{{\cal I}}
\nc{\cJ}{{\cal J}} \nc{\cK}{{\cal K}} \nc{\cL}{{\cal L}}
\nc{\cM}{{\cal M}} \nc{\cN}{{\cal N}} \nc{\cO}{{\cal O}}
\nc{\cP}{{\cal P}} \nc{\cQ}{{\cal Q}} \nc{\cR}{{\cal R}}
\nc{\cS}{{\cal S}} \nc{\cT}{{\cal T}} \nc{\cU}{{\cal U}}
\nc{\cV}{{\cal V}} \nc{\cW}{{\cal W}} \nc{\cX}{{\cal X}}
\nc{\cZ}{{\cal Z}}

\nc{\hA}{{\hat{A}}} \nc{\hB}{{\hat{B}}} \nc{\hC}{{\hat{C}}}
\nc{\hD}{{\hat{D}}} \nc{\hE}{{\hat{E}}} \nc{\hF}{{\hat{F}}}
\nc{\hG}{{\hat{G}}} \nc{\hH}{{\hat{H}}} \nc{\hI}{{\hat{I}}}
\nc{\hJ}{{\hat{J}}} \nc{\hK}{{\hat{K}}} \nc{\hL}{{\hat{L}}}
\nc{\hM}{{\hat{M}}} \nc{\hN}{{\hat{N}}} \nc{\hO}{{\hat{O}}}
\nc{\hP}{{\hat{P}}} \nc{\hR}{{\hat{R}}} \nc{\hS}{{\hat{S}}}
\nc{\hT}{{\hat{T}}} \nc{\hU}{{\hat{U}}} \nc{\hV}{{\hat{V}}}
\nc{\hW}{{\hat{W}}} \nc{\hX}{{\hat{X}}} \nc{\hZ}{{\hat{Z}}}

\nc{\hn}{{\hat{n}}}

\def\w{\mathop{\rm W}}

\newcommand{\bra}[1]{\langle#1|}
\newcommand{\ket}[1]{|#1\rangle}

\newcommand{\norm}[1]{\lVert#1\rVert}

\def\Dbar{\leavevmode\lower.6ex\hbox to 0pt
	{\hskip-.23ex\accent"16\hss}D}
\begin{document}
	\title{The entanglement criteria based on equiangular tight frames}
	
	\author{Xian Shi}\email[]
	{shixian01@gmail.com}
	\affiliation{College of Information Science and Technology,
		Beijing University of Chemical Technology, Beijing 100029, China}

	%\author{Yi Shen}\email[]
	%{yishen@buaa.edu.cn}
	%\affiliation{School of Mathematics and Systems Science, Beihang University, Beijing 100191, China}
	%
	%\author{Yize Sun}
	%\affiliation{School of Mathematics and Systems Science, Beihang University, Beijing 100191, China}
	
	%\author{Lijun Zhao}
	%\affiliation{School of Mathematics and Systems Science, Beihang University, Beijing 100191, China}
	
	%\author{Yumin Guo}
	%\affiliation{School of Mathematical Sciences, Capital Normal University, Beijing 100048, China}
	
	\date{\today}
	
	
	\begin{abstract}
 Finite tight frames play an important role in miscellaneous areas, including quantum information theory. Here we apply a class of tight frames, equiangular tight frames, to address the problem of detecting the entanglement of bipartite states. Here we derive some entanglement criteria based on positive operator-valued measurements built from equiangular tight frames. We also present a class of entanglement witnesses based on the equiangular tight frames. At last, we generalize the entanglement criterion for bipartite systems to multipartite systems.
 	\end{abstract}
 \maketitle
	\section{Introduction}
\indent	Entanglement is one of the essential features in quantum mechanics when compared with classical physics \cite{horodecki2009quantum,plenio2014introduction}. It plays key roles in quantum information processing, such as quantum cryptography \cite{ekert1991quantum}, teleportation \cite{bennett1993teleporting}, and superdense coding \cite{bennett1992communication}. \par
 One of the most important problems in quantum information theory is distinguishing separable and entangled states. If a quantum state $\r_{AB}$ can be written as a convex combination of product states,
 \begin{align*}
 \r_{AB}=\sum_{i=1}^n p_i\rho_A^i\otimes\r_B^i,
 \end{align*}here $\sum_i p_i=1,p_i>0,i=1,2,\cdots, n,$ $\rho_A^i$ and $\rho_B^i$ are states of subsystems $A$ and $B$, respectively, then it is separable. Otherwise, it is entangled. The above problem is completely solved for $2\otimes2$ and $2\otimes3$ systems by the Peres-Horodecki criterion: a bipartite state $\r_{AB}$ is separable if and only if it is positive partial transpose (PPT), $i.$$e.$, $(id\otimes T)(\rho_{AB})\ge 0$ \cite{peres1996separability}. However, the problem is NP-hard for arbitrary dimensional systems \cite{gurvits2003classical}.  In the past twenty years, there have been several other prominent criteria. The computable cross norm or realignment criterion (CCNR) criterion is proposed by Rudolph \cite{rudolph2005further} and Chen and Wu \cite{chen2003matrix}. In 2006, the authors proposed the local uncertainty relations (LURs) and showed that the LURs is stronger than the CCNR criterion \cite{guhne2006entanglement}. In 2007, the author proposed a criterion that is based on Bloch representations \cite{de}. Then Zhang $et$ $al.$ presented the enhanced realignment criterion \cite{zhang2008entanglement}. In 2015, the authors proposed an improved CCNR criterion which they showed is stronger than the CCNR criterion \cite{shen2015separability}. Subsequently, the authors showed that the detection power of the criteria is equivalent to the enhanced realignment criterion \cite{sarbicki2020enhanced}. Jivulescu $et$ $al.$ proposed a class of entanglement criteria via projective tensor norms \cite{Jivulescu2022}. In 2022, Yan $et$ $al.$ proposed several entanglement detection criteria using quantum designs \cite{Yan2021}.  Recently, we proposed a family of separability criteria and presented lower bounds of concurrence and the convex-roof extended negativity for arbitrary dimensional systems \cite{shi2023family}. \par 
Compared with the above methods to detect entanglement for a bipartite system, the criteria for detecting entanglement based on quantum measurements are more easily implemented experimentally. In \cite{spengler2012entanglement}, the authors provided criteria based on mutually unbiased bases (MUBs).  In 2018,  Shang $et$ $al.$ presented a sufficient condition for the separability of a bipartite state based on the symmetric informationally complete (SIC) positive operator-valued measures (POVM) \cite{shang2018enhanced}. \par 
Another method to detect the entanglement of a bipartite mixed state is based on a tool, positive map. The positive map can present a sufficient and necessary condition on the separability of a bipartite state, $i.$$e.$, a quantum state $\rho_{AB}$ is separable if and only if $(I\otimes\Psi)[\r]\ge0$ for any positive map $\Psi(\cdot)$ \cite{horodecki2001separability}. A Hermite operator $W$ on a bipartite system $\mathcal{H}_{AB}$ is an entanglement witness if $W$ is not positive and $\bra{\psi_1\otimes\psi_2}W\ket{\psi_1\otimes\psi_2}\ge 0,$ here $\ket{\psi_1}$ and $\ket{\psi_2}$ are arbitrary pure states in $\mathcal{H}_A$ and $\mathcal{H}_B,$ respectively \cite{lewenstein2001characterization,guhne2009entanglement,chruscinski2014entanglement}. An entanglement witness $W$ is related to a positive but not completely positive map $\Psi$ by the following 
\begin{align}
W=\sum_{i,j=0}^{d-1}\ket{i}\bra{j}\otimes \Psi(\ket{i}\bra{j}),\label{l0}
\end{align}
here $\{\ket{k}\}_{k=0}^{d-1}$ is a set of orthonormal basis in $\mathcal{H}_d.$ Recently, based on the SIC POVM and MUBs, some classes of positive maps and witness have been proposed \cite{chruscinski2018entanglement,siudzinska2022indecomposability}.\par 
 Equiangular tight frames(ETFs) are meaningful subject with a history of at least fifty years \cite{welch1974lower,lemmens1991equiangular,tropp2005complex,fickus2012steiner}. This concept plays a key role in some communication tasks \cite{strohmer2003grassmannian} and sparse approximation \cite{tropp2004greed}. Also, it is closely related to other mathematical subjects, such as graph theory \cite{seidel1991two,waldron2009construction}. The authors in \cite{perez2022mutually} recently generalized the concept to mutually unbiased frames. In a sense, ETFs are tightly relevant to the MUBs \cite{ivonovic1981geometrical,wootters1989optimal} and SIC POVMs \cite{renes2004symmetric}. Due to the same fidelity between any different vectors in a frame, the results on the ETF maybe helpful to study MUBs. Next the maximal sets of complex ETFs can provide SIC POVM. Besides, the existence of SIC POVMs in any dimensional system is an open problem in quantum information theory \cite{horodecki2022five}. Comparing with SIC POVMs, the ETFs without the constraints of the maximal elements are easier to build. In \cite{f2020}, the author present a method to generate new ETFs from the existing ones.  These may motivate us to study ETFs in arbitrary dimensional systems, including their applications. Here we will address the issue of entanglement detection by applying the ETFs.\par 
\indent  This paper is organized as follows.  In section \uppercase\expandafter{\romannumeral2}, we present some preliminary knowledge of this article. In section \uppercase\expandafter{\romannumeral3}, we present our main results. First we present some entanglement criteria for states on bipartite systems,then we offer a class of positive maps and entanglement witness based on ETFs. At last, we proposed some criteria of full separability for a tripartite systems, In section \uppercase\expandafter{\romannumeral5}, we end with a summary.
\section{Preliminaries}
\indent In this section, we recall some knowledge on equiangular tight frames needed in this manuscript. Here we assume all the frames are in complex domain. Readers who are interested in ETFs can refer to \cite{sustik2007existence,fickus2012steiner,waldron2009construction,fer2014}. \par 
Assume $\mathcal{H}_d$ is a Hilbert space with dimension $d$. A set of $n\ge d$ unit vectors $\mathcal{F}=\{\ket{\phi_i}\}_{i=1}^n$ is a frame if there exists $0<s_0<s_1<\infty$ such that 
\begin{align}
s_0\le \sum_{i=1}^n|\bra{\phi_i}\psi\rangle|^2\le s_1, \hspace{3mm} \forall \ket{\psi}\in \mathcal{H}_d.\label{l1}
\end{align}
From the definition above, we have $s_0$ and $s_1$ can be seen as the minimal and maximal eigenvalues of the following operator, respectively,
\begin{align}
S=\sum_{i=1}^n\ket{\phi_i}\bra{\phi_i}.\label{l2}
\end{align}
When $S=bI$ for some positive number $b$, $\{\phi_i\}_{i=1}^n$ is a tight frame. In other words, when $\{\phi_i\}_{i=1}^n$ is a tight frame, $s_0$ and $s_1$ in (\ref{l1}) are equal to $b.$ As $\ket{\phi_i}$ are all unit vectors, $b=\frac{n}{d}.$ And when $n=d,$ the tight frames can be seen as an orthonormal base. Next, a tight frame is called equiangular when there exists $c>0$ such that 
\begin{align}
|\bra{\phi_i}\phi_j\rangle|^2=c,\hspace{3mm} i\ne j, \label{l3}
\end{align}
and
\begin{align*}
b=\frac{n}{d},\hspace{3mm} c=\frac{n-d}{(n-1)d}.
\end{align*}
If there exists an ETF with $n$ elements in $d$ dimensions, then $n\le d^2.$ When $n=d^2,$ $c=|\bra{\phi_i}\phi_j\rangle|^2=\frac{1}{d+1},$ the ETF turns into a SIC POVM. \par 
Due to the definition of a tight frame, $\{\ket{\phi_i}\}$ satisfy the following property,  $$\frac{d}{n}\sum_{i=1}^n\ket{\phi_i}\bra{\phi_i}=I.$$
 Hence an ETF can be seen as a POVM $\mathcal{E}=\{E_i=\frac{d}{n}\ket{\phi_i}\bra{\phi_i}\}_{i=1}^n.$ Then for a quantum state $\r,$ the probability of the $i$-th outcome is 
 \begin{align*}
 p_i(\mathcal{E},\rho)=\frac{d}{n}\bra{\phi_i}\rho\ket{\phi_i}.
 \end{align*}\par 
Recently, Rastegin derived fine-grained and entropic uncertainty relations for the measurements based on the ETFs \cite{rastegin2023}. There the author showed that 
\begin{align}
I(\mathcal{E},\r)=\sum_i p_i(\mathcal{E},\r)^2\le \frac{bc+(1-c)tr\rho^2}{b^2},\label{l4}
\end{align} 
here $b=\frac{n}{d},$ $c=\frac{n-d}{(n-1)d}$.
\section{Entanglement Criterion Based on ETFs}
\indent In this subsection, based on the ETFs, we present some entanglement criteria for arbitrary dimensional bipartite states.\par 
Assume $\r$ is a bipartite state on $\mathcal{H}=\mathcal{H}_A\otimes\mathcal{H}_B$ with dimension $d_A=d_B=d,$ and denote $\mathcal{E}^A=\{E_i^A=\frac{d}{n}\ket{\phi_i}\bra{\phi_i}\}_{i=1}^n$ and $\mathcal{E}^B=\{E_j^B=\frac{d}{n}\ket{\psi_j}\bra{\psi_j}\}_{j=1}^n$ as two POVMs generated by ETFs $\{\ket{\phi_i}\}_{i=1}^n$ and $\{\ket{\psi_j}\}_{j=1}^n$ with $c_A$ and $c_B$, respectively. Let $M(\mathcal{E}_A,\mathcal{E}_B,\rho_{AB})_{d\times d}$ be a matrix with its element
\begin{align*}
[M(\mathcal{E}_A,\mathcal{E}_B,\rho_{AB})]_{ij}=tr[(E_i^A\otimes E_j^B)\rho_{AB}],
\end{align*}
based on this matrix, we will first present an entanglement criterion for bipartite systems via ETFs.
\begin{Theorem}\label{t1}
	Assume $\rho_{AB}$ is separable, then 
	\begin{align*}
	\norm{M(\mathcal{E}_A,\mathcal{E}_B,\rho_{AB})}_1\le \sqrt{\frac{(b_Ac_A+(1-c_A))(b_Bc_B+(1-c_B))}{b_A^2b_B^2}},
	\end{align*}
	otherwise, $\r_{AB}$ is entangled. 
\end{Theorem}
Here we place the proof of Theorem \ref{t1} in Sec. \ref{app}.
\par 
When $b_A=b_B=d,$ $c_A=c_B=\frac{1}{d+1},$ that is, $\{E_i^A\}$ and $\{E_j^B\}$ turn into the normalized SIC POVM, then Theorem \ref{t1} turns into $\norm{M(\mathcal{E}_A,\mathcal{E}_B,\rho_{AB})}_1\le \frac{2}{d(d+1)},$ which is the ESIC in \cite{shang2018enhanced}. \par 
Next we present some examples by the above theorem. 
\begin{example}
Consider the following $d\times d$ dimensional isotropic states
\begin{align*}
\rho_{AB}=\frac{1-p}{d^2}I_{d^2}+p\ket{\psi_d^{+}}\bra{\psi_d^{+}}, \hspace{3mm} p\in[0,1],
\end{align*}
here $\ket{\psi^{+}_d}=\frac{1}{\sqrt{d}}\sum_i\ket{ii}$. The isotropic states $\r_{AB}$ is entangled if and only if $p>\frac{1}{d+1}$ \cite{horodecki1999reduction}.\par 
Let $\{E_i^A=\frac{d}{n}\ket{\phi_i}\bra{\phi_i}\}_{i=1}^n$ be a POVM generated by a ETF $\{\ket{\phi_i}\}_{i=1}^n$ with $\sum_i \ket{\phi_i}\bra{\phi_i}=b I,$ let $\{E_i^B=\frac{d}{n}\ket{\phi^{*}_i}\bra{\phi^{*}_i}\},$ here $\ket{\phi_i^{*}}$ is complex conjugate to $\ket{\phi_i}$. As 
\begin{align*}
\sum_i\ket{\phi^{*}_i}\bra{\phi_i^{*}}=\sum_i\overline{\ket{\phi_i}\bra{\phi_i}}=bI,\\
|\bra{\phi^{*}_i}\phi^{*}_j\rangle|^2=|\bra{\phi_i}\phi_j\rangle|^2=c\hspace{3mm}\forall i\ne j,
\end{align*} 
that is, $\{\ket{\phi_i^{*}}\}_{i=1}^n$ is also a ETF. Then we have 
\begin{align}
[M(\mathcal{E}_A,\mathcal{E}_B,\rho_{AB})]_{k,k}=&\frac{pd+1-p}{n^2}\hspace{5mm} \nonumber\\
[M(\mathcal{E}_A,\mathcal{E}_B,\rho_{AB})]_{k,l}=&\frac{pdc+1-p}{n^2}\hspace{5mm}k\ne l
\end{align} 
through computation,  $$\norm{M(\mathcal{E}_A,\mathcal{E}_B,\rho_{AB})}_1=\frac{1-p+pd}{n},$$
hence when choosing an ETF with $n$ elements, the above criterion can detect entanglement when
\begin{align*}
p> \frac{d-1}{n-1},
\end{align*}
when $n=d^2,$ $p>\frac{1}{d+1},$ that is, we can identify the entanglement of all the isotropic states in arbitrary dimensional systems.
\end{example}
\begin{example}
Consider the following states of bound entangled states $$\s(x,p)=p\rho_x+\frac{(1-p)I}{9},$$
here $p\in[0,1],$ $\rho_x$ is a bound entangled state with $x\in(0,1)$ proposed by Horodecki \cite{HORODECKI1997},
\begin{align*}
\rho_x=\frac{1}{1+8x}\begin{pmatrix}
x&0&0&0&x&0&0&0&x\\
0&x&0&0&0&0&0&0&0\\
0&0&x&0&0&0&0&0&0\\
0&0&0&x&0&0&0&0&0\\
x&0&0&0&x&0&0&0&x\\
0&0&0&0&0&x&0&0&0\\
0&0&0&0&0&0&\frac{1+x}{2}&0&\frac{\sqrt{1-x^2}}{2}\\
0&0&0&0&0&0&0&x&0\\
x&0&0&0&x&0&\frac{\sqrt{1-x^2}}{2}&0&\frac{1+x}{2}
\end{pmatrix}.
\end{align*}
Let \begin{align}
\ket{\phi_1}=&\frac{1}{\sqrt{2}}\begin{pmatrix}
0\\1\\-1
\end{pmatrix},\hspace{1mm}\ket{\phi_2}=\frac{1}{\sqrt{2}}\begin{pmatrix}
-1\\0\\1
\end{pmatrix},\hspace{1mm}\ket{\phi_3}=\frac{1}{\sqrt{2}}\begin{pmatrix}
1\\-1\\0
\end{pmatrix},\nonumber\\
&\ket{\phi_4}=\frac{1}{\sqrt{2}}\begin{pmatrix}
0\\w\\-w^2
\end{pmatrix},
\hspace{5mm}\ket{\phi_5}=\frac{1}{\sqrt{2}}\begin{pmatrix}
-1\\0\\w^2
\end{pmatrix},\nonumber\\&\ket{\phi_6}=\frac{1}{\sqrt{2}}\begin{pmatrix}
1\\-w\\0
\end{pmatrix},\hspace{5mm}
\ket{\phi_7}=\frac{1}{\sqrt{2}}\begin{pmatrix}
0\\w^2\\-w
\end{pmatrix},\nonumber\\&\ket{\phi_8}=\frac{1}{\sqrt{2}}\begin{pmatrix}
-1\\0\\w
\end{pmatrix},\hspace{5mm}\ket{\phi_9}=\frac{1}{\sqrt{2}}\begin{pmatrix}
1\\-w^2\\0
\end{pmatrix}, \label{e1}
\end{align}
here $w=e^{\frac{2}{3}\pi i}$, and the nine states $\{\ket{\phi_i}|i=1,2,\cdots,9\}$ make up a SIC POVM \cite{stacey2016sic}. We take POVMs $\{E_i^A=\frac{1}{3}\ket{\phi_i}\bra{\phi_i}\}_{i=1}^9$ and $\{E_i^B=\frac{1}{3}\ket{\phi_i^{*}}\bra{\phi_i^{*}}\}_{i=1}^9$ for systems A and B, respectively.\par 

Next we take another ETF with $n=7$ \cite{sustik2007existence},
\begin{align*}
\ket{\phi_1}=&\frac{1}{\sqrt{3}}\begin{pmatrix}
1\\1\\1
\end{pmatrix},\hspace{2mm}\ket{\phi_2}=\frac{1}{\sqrt{3}}\begin{pmatrix}
1\\e^{\frac{2\pi i}{7}}\\e^{\frac{6\pi i}{7}}
\end{pmatrix},\hspace{2mm}
\ket{\phi_3}=\frac{1}{\sqrt{3}}\begin{pmatrix}
1\\e^{\frac{4\pi i}{7}}\\e^{\frac{12\pi i}{7}}
\end{pmatrix},\\
&\ket{\phi_4}=\frac{1}{\sqrt{3}}\begin{pmatrix}
1\\e^{\frac{6\pi i}{7}}\\e^{\frac{4\pi i}{7}}
\end{pmatrix},\hspace{5mm}
\ket{\phi_5}=\frac{1}{\sqrt{3}}\begin{pmatrix}
1\\e^{\frac{8\pi i}{7}}\\e^{\frac{10\pi i}{7}}
\end{pmatrix},\nonumber\\&\ket{\phi_6}=\frac{1}{\sqrt{3}}\begin{pmatrix}
1\\e^{\frac{10\pi i}{7}}\\e^{\frac{2\pi i}{7}}
\end{pmatrix},\hspace{5mm}\ket{\phi_7}=\frac{1}{\sqrt{3}}\begin{pmatrix}
1\\e^{\frac{12\pi i}{7}}\\e^{\frac{8\pi i}{7}}
\end{pmatrix},\nonumber
\end{align*}
here $\a=\frac{1+\sqrt{5}}{2}.$\par
\indent  In Fig. \ref{f}, we present the effects on detecting the entanglement of $\s(x,p)$ based on the ETFs when $n=7$ and $n=9$. From the figures, we can see that the detection power of entanglement based on the ETF when $n=9$ is better than the scenario when $n=7.$
\begin{figure*}
	\subfigure[n=7]{\label{f1}
	\includegraphics[scale=0.6]{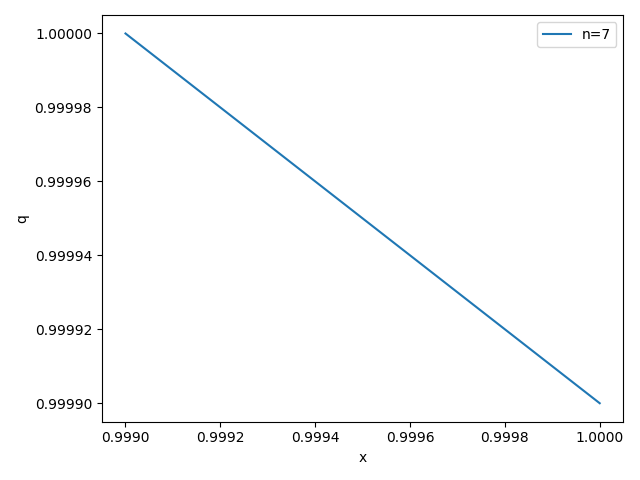}
}
	\subfigure[n=9]{\label{f2}
	\includegraphics[scale=0.6]{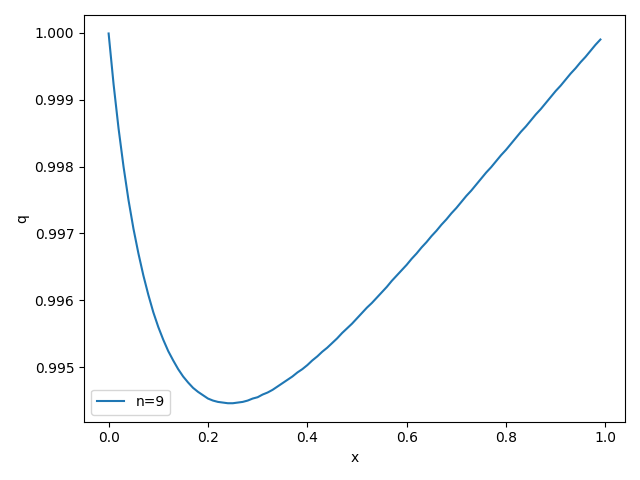}
}
	\caption{Comparison with the detection power of entanglement between the ETFs when $n=7$ and $n=9.$ From the above figures, we see that the detection power of entannglement based on the ETF when $n=9$ is better than the case when $n=7.$} \label{f}
\end{figure*}
\end{example}
\section{Positive maps and entanglement witnesses}
In this section, we will apply an ETF to present the positive map on arbitrary dimensional systems. Then based on the positive map, we will give some entanglement witnesses.\par 
Assume $\{\ket{\phi_i}\}_{i=1}^n$ constitute an ETF in a Hilbert space $\mathcal{H}_d$, then $\frac{d}{n}\sum_{i=1}^n\ket{\phi_i}\bra{\phi_i}=I$, that is, $\mathcal{P}=\{P_i=\frac{d}{n}\ket{\phi_i}\bra{\phi_i}\}_{i=1}^n$ constitutes a POVM on $\mathcal{H}_d$. Let $\mathcal{O}$ be an orthogonal rotation in $\mathbb{R}^{n}$ satisying the following properties,
\begin{itemize}
	\item[(i)] \begin{align}\sum_k\mathcal{O}_{kl}=1,\hspace{3mm}\forall l\in\{1,2,\cdots, n\}\label{i1}\end{align}
	\item[(ii)] \begin{align}\sum_k \mathcal{O}_{kl}\mathcal{O}_{km}=\delta_{lm}, \label{i2}\end{align}
\end{itemize}
here $\delta_{lm}$ is the Kronecker symbol. Next we consider the following trace-preserving map $\Phi(\cdot),$
\begin{widetext}
\begin{align}
\Phi(X)=\sqrt{\frac{d(d-1)}{n(n-1)}}\left[\frac{I_d}{d}tr(X)-h\sum_{k,l=1}^n\mathcal{O}_{kl}tr[(X-\frac{I_d}{d}tr(X))P_l]P_k\right],\label{l7}
\end{align}
\end{widetext}
here $h=\sqrt{\frac{n^3(n-1)}{d^3(d-1)^3}},$ $P_l$ and $P_k$ takes over all the elements in the POVM $\mathcal{P}.$ 

\begin{Theorem}\label{t3}
The trace-preserving map $\Phi(\cdot)$ defined in (\ref{l7}) is positive.
\end{Theorem}
\par Here we present the proof in Sec. \ref{app}.
\par One of the important applications of positive maps is to detect the entanglement of 
bipartite systems. Based on (\ref{l0}), we can provide a class of entanglement witnesses for states on systems $\mathcal{H}_d\otimes\mathcal{H}_d$ relative to $(\ref{l7}):$
\begin{align}
W=\sqrt{\frac{d(d-1)}{n(n-1)}}\left[\frac{1}{d}I_d\otimes I_d-h\sum_{k,l}\mathcal{O}_{kl} (\overline{\Psi}(P_l))^T\otimes P_k\right],
\end{align}
here $\overline{\Psi}(P_l)=P_l-\frac{I_d}{d}tr P_l.$\par 
In the next section, we generalize the methods to detect the entanglement for bipartite systems based on the ETFs to indentify the genuine entanglement for tripartite systems.
\section{Separability criteria for tripartite systems}
\indent Given $\rho_{ABC}$ a tripartite mixed state on $\mathcal{H}_{d}\otimes\mathcal{H}_d\otimes\mathcal{H}_d,$ and assume the POVM $\mathcal{E}_A$, $\mathcal{E}_B$ and $\mathcal{E}_C$ are generated by $\{\ket{\phi_i}_A\}_{i=1}^{n_A}$, $\{\ket{\theta_j}_B\}_{j=1}^{n_B}$, and $\{\ket{\omega_k}_C\}_{k=1}^{n_C},$ respectively. That is, 
\begin{align*}
\frac{n_A}{d}\sum_{i=1}^{n_A}\ket{\phi_i}\bra{\phi_i}=I_A,\\
\frac{n_B}{d}\sum_{j=1}^{n_B}\ket{\theta_j}\bra{\theta_j}=I_B,\\\frac{n_C}{d}\sum_{k=1}^{n_C}\ket{\omega_k}\bra{\omega_k}=I_C,
\end{align*}
Next we recall a hypermatrix $\mathcal{M}_{n_A\times n_B\times n_C}=[m_{ijk}]$ on $\r_{ABC}$ and its \emph{Frobenius} norm, here the elements $m_{ijk}$ in $\mathcal{M}$ are defined as
\begin{align}
m_{ijk}= tr (\rho_{ABC}(P_i^A\otimes P_j^B\otimes P_k^C) ),
\end{align}
where  $i\in \{1,2,\cdots,n_A\},j\in\{1,2,\cdots,n_B\}$ and $k\in\{1,2,\cdots,n_C\}.$ The \emph{Frobenius} norm of $\mathcal{M}_{n_A\times n_B\times n_C}$ is defined as 
\begin{align}
\norm{\mathcal{M}}_{F}:=\sqrt{\sum_{i_1=1}^{n_A}\sum_{i_2=1}^{n_B}\sum_{i_3=1}^{n_C}|m_{i_1i_2i_3}|^2}
\end{align}
Then we apply the \emph{Frobenius} norm to present a criterion on the fully separability criterion of a tripartite state.
\begin{Theorem}\label{t4}
	Assume $\r_{ABC}$ is a tripartite fully separable state, then
	\begin{align}
	&\norm{\mathcal{M}_{\rho}}_{F}\nonumber\\
	\le&\frac{\sqrt{(b_Ac_A+1-c_A)(b_Bc_B+1-c_B)(b_Cc_C+1-c_C)}}{b_Ab_Bb_C}.
	\end{align}
	Here $b_i=\frac{n_i}{d},c_i=\frac{n_i-d}{n_id-d},$ $i$ is on behalf of $A,$ $B$ or $C.$
\end{Theorem}
We present the proof of Theorem \ref{t4} in Sec. \ref{app}.\par
At last, we will consider the detection of full separability for a tripartite mixed state by studying its biseparability. When $\r_{ABC}$ is a fully separable mixed state on $\mathcal{H}_A\otimes\mathcal{H}_B\otimes\mathcal{H}_C$, then $\r_{ABC}$ is biseparable in terms of the partition $A|BC,$ $B|AC$ and $C|AB.$ Let
\begin{align}
E^{\mathbf{\underline{a}bc}}(\r)=E_1^{(1)}\oplus E_1^{(2)}\oplus \cdots\oplus E_1^{(n)},
\end{align}
with $[E_1^{(i)}]_{jk}=c_{ijk},$ $i=1,2,\cdots n.$ 
\begin{align}
E^{\mathbf{\underline{b}ac}}(\r)=E_2^{(1)}\oplus E_2^{(2)}\oplus \cdots\oplus E_2^{(n)},
\end{align}
with $[E_2^{(j)}]_{ik}=c_{ijk},$ $j=1,2,\cdots n.$ 
\begin{align}
E^{\mathbf{\underline{c}ab}}(\r)=E_3^{(1)}\oplus E_3^{(2)}\oplus \cdots\oplus E_3^{(n)},
\end{align}
with $[E_3^{(k)}]_{ij}=c_{ijk},$ $k=1,2,\cdots n.$ 
\begin{Theorem}\label{t5}
	Assume $\r_{ABC}$ is  fully separable state on $\mathcal{H}_d\otimes\mathcal{H}_d\otimes\mathcal{H}_d$, then we have 
	\begin{align}
\norm{E^{\mathbf{\underline{a}bc}}(\rho)}_1\le \sqrt{\frac{(n_B-2d+d^2)(n_C-2d+d^2)}{n_Bn_C(n_B-1)(n_C-1)}},\nonumber\\
\norm{E^{\mathbf{\underline{b}ac}}(\rho)}_1\le  \sqrt{\frac{(n_A-2d+d^2)(n_C-2d+d^2)}{n_An_C(n_A-1)(n_C-1)}},\nonumber\\
\norm{E^{\mathbf{\underline{c}ab}}(\rho)}_1\le \sqrt{\frac{(n_B-2d+d^2)(n_A-2d+d^2)}{n_An_B(n_B-1)(n_A-1)}}.
	\end{align}
\end{Theorem}
Here we present the proof of this theorem in Sec. \ref{app}
\begin{example}
	Consider the following three qubit mixed state
	\begin{align}
	\r(x)=\frac{x}{27}I_3^{\otimes3}+(1-x)\ket{\phi}\bra{\phi}, \hspace{3mm} x\in[0,1].
	\end{align}
	Here $\ket{\phi}=\frac{1}{\sqrt{6}}(\ket{123}-\ket{132}+\ket{231}-\ket{213}+\ket{312}-\ket{321})$, it was proposed in \cite{ou2007violation} to show that higher dimensional systems may not satisfy the monogamy relations. As $\ket{\phi}$ is asymmetrical, $\r(x)$ stays the same when swapping any two partites. When applying the POVM generated by ETF $(\ref{e1})$ to all parties, we have 
	\begin{align*}
	\norm{E^{\underline{a}bc}(\rho(x))}_1=\norm{E^{\underline{b}ac}(\rho(x))}_1=\norm{E^{\underline{c}ab}(\rho(x))}_1,
	\end{align*}
	In Fig. \ref{a}, we plot the regions when $\r(x)$ is fully separable based on Theorem \ref{t5}.
	\begin{figure}
		\centering
		\includegraphics[width=85mm]{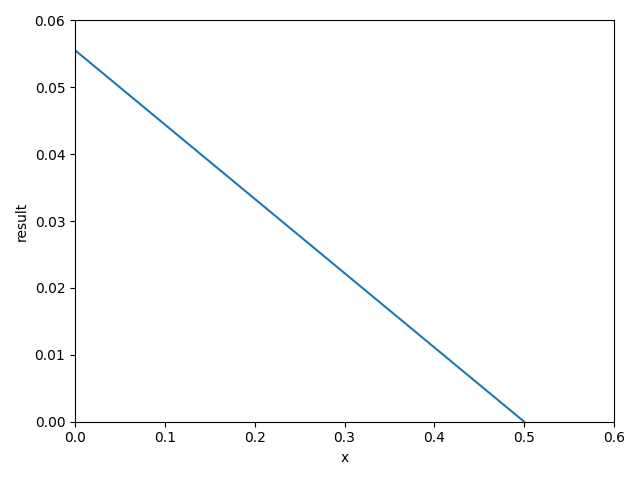}
		\caption{Entanglement detection of $\r(x)$.  }
		\label{a}
	\end{figure}
\end{example}
\section{Summary}
\indent In this paper, we have addressed the detection of bipartite and multipartite entanglement by applying the positive operator valued measurements based on the equiangular tight frames (ETF). The measurements are meaningful not only due to the beautiful structure of itself, but also due to its highly connections with some critical elements in quantum information theory. Here we presented an entanglement criterion based on the (ETF) for mixed states on arbitrary dimensional systems.  And also we presented a class of positive maps based on the ETF, which can construct a class of entanglement witnesses directly. At last, we generalized the bipartite entanglement criterion to tripartite entangled systems.  Due to the importance of ETF, we hope the above results could shed some light on related studies in quantum information theory.
  \section{Acknowledgement}
  X. S. was supported by the Fundamental Research Funds for the Central Universities (Grant No. ZY2306), and Funds of College of Information Science and Technology, Beijing University of Chemical Technology (Grant No. 0104/11170044115).
  \bibliographystyle{IEEEtran}
\bibliography{ref}
\section{Appendix }\label{app}
\textbf{Theorem \ref{t1}}  \emph{	Assume $\rho_{AB}$ is separable, then 
\begin{align*}
\norm{M(\mathcal{E}_A,\mathcal{E}_B.\rho_{AB})}_1\le \sqrt{\frac{(b_Ac_A+(1-c_A))(b_Bc_B+(1-c_B))}{b_A^2b_B^2}},
\end{align*}
otherwise, $\r_{AB}$ is entangled. }

\begin{proof}
	Assume $\r_{AB}$ is a product state $\r_{AB}=\r_A\otimes\r_B,$ then
	\begin{align*}
	M(\mathcal{E}_A,\mathcal{E}_B,\rho_{AB})=\begin{pmatrix}
	e^A_1\\e^A_2\\\vdots\\e^A_{n}
	\end{pmatrix}\begin{pmatrix}
	e^B_1&e^B_2&\cdots&e^B_n
	\end{pmatrix},
	\end{align*} 
	here $e_i^A=tr E_i^A\r_A,$  $i=1,2,\cdots,n_A.$ And $e_j^B=tr E_j^B\r_B,$ $ j=1,2,\cdots,n_B.$ Then we have 
	\begin{align}
	&\norm{M(\mathcal{E}_A,\mathcal{E}_B,\rho_{AB})}_{1}\nonumber\\
	=&\sqrt{I(\mathcal{E}^A,\r_A)I(\mathcal{E}^B,\r_B)}\nonumber\\
	\le&\sqrt{\frac{(b_Ac_A+(1-c_A)tr\rho_A^2)(b_Bc_B+(1-c_B)tr\rho_B^2)}{b_A^2b_B^2}}\nonumber\\
	\le&\sqrt{\frac{(b_Ac_A+(1-c_A))(b_Bc_B+(1-c_B))}{b_A^2b_B^2}}.\label{ll}
	\end{align}
	Here the first inequality is due to (\ref{l4}), and the second inequality is due to $tr\rho_A^2,tr\rho_B^2\le 1$. \par 
	At last, when $\rho_{AB}=\sum_k p_k\rho^A_k\otimes\rho^B_k,$ by utilizing the convexity of $\norm{\cdot}_1$, we have
	\begin{align*}
	&\norm{M(\mathcal{E}_A,\mathcal{E}_B,\rho_{AB})}_1\nonumber\\
	\le& \sum_kp_k\norm{M(\mathcal{E}_A,\mathcal{E}_B,\rho^A_i\otimes\rho^B_i)}\nonumber\\
	\le&\sqrt{\frac{(b_Ac_A+(1-c_A))(b_Bc_B+(1-c_B))}{b_A^2b_B^2}}.
	\end{align*}
	We finish the proof.
\end{proof}

\textbf{Theorem \ref{t3}}
\emph{The trace-preserving map $\Phi(\cdot)$ defined in (\ref{l7}) is positive.}\\
\begin{proof}
	As the coefficient $\sqrt{\frac{d(d-1)}{n(n-1)}}$ is positive, the positivity of $\Psi(X)=\sqrt{\frac{n(n-1)}{d(d-1)}}\Phi(X)$ is equivalent to the positivity of $\Phi(\cdot).$
	Next let $P=\ket{\phi}\bra{\phi},$ we will prove $tr[\Psi(P)]^2\le \frac{1}{d-1},$ which is a sufficient condition when $\Psi(\cdot)$ is positive \cite{chruscinski2018entanglement}. Let $\overline{\Psi}(X)=X-\frac{tr X}{d}I,$ 
	
	\begin{align}
	&tr[\Psi(P)]^2\nonumber\\=&Tr\big[\frac{I_d}{d^2}-\frac{2h}{d}\sum_{k,l=1}^{n}\mathcal{O}_{kl}tr(\overline{\Psi}(P)P_l)P_k\nonumber\\
	+&h^2\sum_{k,l,s,t=1}^{n}\mathcal{O}_{kl}\mathcal{O}_{st}tr[\overline{\Psi}(P)P_l]P_ktr[\overline{\Psi}(P)P_t]P_s\big]\nonumber\\
	=&\frac{1}{d}-\frac{2h}{d}\sum_{k,l=1}^{n}\mathcal{O}_{kl}tr(\overline{\Psi}(P)P_l)tr(P_k)\nonumber\\
	+&h^2\sum_{k,l,s,t=1}^{n}\mathcal{O}_{kl}\mathcal{O}_{st}tr[\overline{\Psi}(P)P_l]tr[\overline{\Psi}(P)P_t]tr(P_kP_s)\nonumber\\
	=&\frac{1}{d}+h^2\sum_{k,l,t=1}^{n}\mathcal{O}_{kl}\mathcal{O}_{kt}tr[\overline{\Psi}(P)P_l]tr[\overline{\Psi}(P)P_t]tr(P_kP_k)\nonumber\\+&h^2\sum_{k\ne s=1}^{n}\sum_{l,t=1}^{n}\mathcal{O}_{kl}\mathcal{O}_{st}tr[\overline{\Psi}(P)P_l]tr[\overline{\Psi}(P)P_t]tr(P_kP_s)\nonumber\\
	=&\frac{1}{d}+\frac{h^2d^2}{n^2}\sum_{k,l,t=1}^{n}\delta_{lt}tr[\overline{\Psi}(P)P_l]tr[\overline{\Psi}(P)P_t]\nonumber\\+&h^2\sum_{k\ne s=1}^{n}\sum_{l,t=1}^{n}\mathcal{O}_{kl}\mathcal{O}_{st}tr[\overline{\Psi}(P)P_l]tr[\overline{\Psi}(P)P_t]tr(P_kP_s)\nonumber\\
	=&\frac{1}{d}+\frac{h^2d^2}{n^2}\sum_{l=1}^{n}[tr(\overline{\Psi}(P)P_l)]^2,\label{l5}
	\end{align}
	here the second equality is due to that $(\ref{i1}),$ the fourth equality is due to that $(\ref{i2})$. Next 
	\begin{align}
	[tr(\overline{\Psi}(P)P_l)]^2=(trPP_l)^2-\frac{2}{n}tr PP_l+\frac{1}{n^2},\label{l6}
	\end{align}
	as \begin{align}
	\sum_l (trPP_l)^2\le   \frac{bc+(1-c)}{b^2}=\frac{n+d^2-2d}{n(n-1)},
	\end{align}
	hence
	\begin{align}
	(\ref{l5})\le& \frac{1}{d}+\frac{h^2d^2}{n^2}\times(\frac{n+d^2-2d}{n(n-1)}-\frac{2}{n}+\frac{n}{n^2})\nonumber\\\le&  \frac{1}{d-1},
	\end{align}
	Then we finish the proof.
\end{proof}

\textbf{Theorem \ref{t4}} Assume $\r_{ABC}$ is a tripartite fully separable state, then
\begin{align}
&\norm{\mathcal{M}_{\rho}}_{F}\nonumber\\
\le&\frac{\sqrt{(b_Ac_A+1-c_A)(b_Bc_B+1-c_B)(b_Cc_C+1-c_C)}}{b_Ab_Bb_C}.
\end{align}
Here $b_i=\frac{n_i}{d},c_i=\frac{n_i-d}{n_id-d},$ $i$ is on behalf of $A,$ $B$ or $C.$

\begin{proof}
	Assume $\r_{ABC}=\r_A\otimes\r_B\otimes\r_C$, 
	\begin{align*}
	m_{i_1i_2i_3}=&tr(\r_A\otimes\r_B\otimes\r_C)(P_{i_1}\otimes P_{i_2}\otimes P_{i_3})\nonumber\\
	=&tr(\r_A P_{i_1})tr(\r_B P_{i_2}) tr(\r_C P_{i_3}),
	\end{align*}
	then
	\begin{align*}
	&\norm{M_{\r}}_{F}\nonumber\\=&\sqrt{\sum_{i_1i_2i_3}|m_{i_1i_2i_3}|^2}\nonumber\\
	=&\sqrt{\sum_{i_1i_2i_3}[tr(\r_A P_{i_1})tr(\r_B P_{i_2}) tr(\r_C P_{i_3})]^2}\nonumber\\
	=&\sqrt{\sum_{i_1}[tr(\r_A P_{i_1})]^2\sum_{i_2}[tr(\r_B P_{i_2})]^2\sum_{i_3}[tr(\r_C P_{i_3})]^2}\nonumber\\
	\le&\frac{\sqrt{(b_Ac_A+1-c_A)(b_Bc_B+1-c_B)(b_Cc_C+1-c_C)}}{b_Ab_Bb_C}.
	\end{align*}
\end{proof}
\textbf{Theorem \ref{t5}} \emph{	Assume $\r_{ABC}$ is  fully separable state on $\mathcal{H}_d\otimes\mathcal{H}_d\otimes\mathcal{H}_d$, then we have 
	\begin{align}
	\norm{E^{\mathbf{\underline{a}bc}}(\rho)}_1\le \sqrt{\frac{(n_B-2d+d^2)(n_C-2d+d^2)}{n_Bn_C(n_B-1)(n_C-1)}},\nonumber\\
	\norm{E^{\mathbf{\underline{b}ac}}(\rho)}_1\le  \sqrt{\frac{(n_A-2d+d^2)(n_C-2d+d^2)}{n_An_C(n_A-1)(n_C-1)}},\nonumber\\
	\norm{E^{\mathbf{\underline{c}ab}}(\rho)}_1\le \sqrt{\frac{(n_B-2d+d^2)(n_A-2d+d^2)}{n_An_B(n_B-1)(n_A-1)}}.
	\end{align}}
\begin{proof}
	As $\r_{ABC}$ is a fully separable state, it can be written as
	\begin{align}
	\r=\sum_i p_i\rho^A_i\otimes\rho^B_i\otimes\rho^C_i,
	\end{align}
	here $\sum_ip_i=1,$ $p_i \in(0,1].$ As $\norm{\cdot}_1$ is a norm, here we only prove when $\rho_{ABC}=\r_A\otimes\r_B\otimes\r_C.$ Under this condition, we have 
	\begin{align*}
	E^{\underline{a}bc}(\rho)=E^\mathbf{a}\otimes e^\mathbf{b}(e^\mathbf{c})^T\nonumber\\
	E^\mathbf{a}=\begin{pmatrix}
	e_1&0&\cdots&0\nonumber\\
	0&e_2&\cdots&0\nonumber\\
	\vdots&\vdots&\ddots& \vdots \\
	0&0&\cdots&e_n
	\end{pmatrix},\nonumber\\
	e^\mathbf{b}=\begin{pmatrix}
	e_1^\mathbf{b}\\e_2^\mathbf{b}\\\vdots \\e_n^\mathbf{b}
	\end{pmatrix},\hspace{3mm}	e^\mathbf{c}=\begin{pmatrix}
	e_1^\mathbf{c}\\e_2^\mathbf{c}\\\vdots \\e_n^\mathbf{c}
	\end{pmatrix},
	\end{align*}
	here 
	\begin{align*}
	e_i^{\mathbf{a}}= tr P_i^A \rho_A, \hspace{3mm} i=1,2,\cdots,n_A,\nonumber\\
	e_j^{\mathbf{b}}= tr P_j^B \rho_B, \hspace{3mm} j=1,2,\cdots,n_B,\nonumber\\
	e_i^{\mathbf{c}}= tr P_i^C \rho_C, \hspace{3mm} i=1,2,\cdots,n_C,
	\end{align*}
	then we have 
	\begin{align}
	&\norm{E^{\underline{a}bc}(\rho^A\otimes\r^B\otimes\r^C)}_1\nonumber\\
	=&\norm{E^a}_1\norm{e^b}_1\norm{e^c}_1\nonumber\\
	=&\sum_i tr(P_i^a\r_A)\norm{e^b}_1\norm{e^c}_1\nonumber\\
	=&\sqrt{\frac{(n_B-2d+d^2)(n_C-2d+d^2)}{n_Bn_C(n_B-1)(n_C-1)}}.
	\end{align}
	The bound of $\norm{E^{\mathbf{\underline{b}ac}}(\r)}_1$ and $\norm{E^{\mathbf{\underline{c}ab}}(\r)}_1$ can be proved similarly.
\end{proof}
\end{document}